\documentclass[12pt]{iopart}
\ifx\pdfoutput\undefined
\usepackage[dvips]{graphicx}
\else
\usepackage[pdftex]{graphicx}
\usepackage{iopams}
\usepackage{epstopdf}
\fi
\usepackage{cite}
\usepackage{hhline}

\newcommand{\halfint}[1]{\ensuremath{\left\lfloor \frac{#1}{2} \right\rfloor}}

\newcommand{\qbinom}[2]{\ensuremath{\left[ {#1 \atop #2} \right]}}
\newcommand{\binom}[2]{\ensuremath{\left( {#1 \atop #2} \right)}}

\newcommand{\invq}{\ensuremath{q^{-1}}}

\renewcommand{\theequation}{\arabic{section}.\arabic{equation}}

 \def\1{\mathsf{1}}
 \def\({\left(}
 \def\){\right)}

\newcommand{\tb}{\tilde{\beta}}
\newcommand{\ta}{\tilde{\alpha}}

\def\la{\langle}
\def\ra{\rangle}

\newcommand{\Cee}{\ensuremath{\mathcal{C}}}

\def\-{{\bf --}} 

\newcommand{\bra}[1]{\ensuremath{\langle #1 |}}
                                                                                
\newcommand{\ket}[1]{\ensuremath{| #1 \rangle}}
                                                                                
\newcommand{\braket}[2]{\ensuremath{\langle #1 | #2 \rangle}}
                                                                                
\newcommand{\braopket}[3]{\ensuremath{\langle #1 | #2 | #3 \rangle}}
                                                                                
\begin{document}


\title{Continued Fractions and the Partially Asymmetric Exclusion Process}
\date{April 2009}

\author{R.\ A.\ Blythe}
\address{SUPA, School of Physics and Astronomy, University
of Edinburgh, Mayfield Road, Edinburgh EH9 3JZ}

\author{W.\  Janke}
\address{Institut f\"ur Theoretische Physik and Centre for Theoretical Sciences (NTZ),
 Universit\"at Leipzig,
 Postfach 100\,920,
04009 Leipzig, Germany}

\author{D.\ A.\ Johnston}
\address{Department of Mathematics and the Maxwell Institute for Mathematical
Sciences,
Heriot-Watt University, Riccarton, Edinburgh EH14 4AS, Scotland}

\author{R.\ Kenna}
\address{Applied Mathematics Research Centre, Coventry University, Coventry, CV1 5FB, England}

\begin{abstract}
We note that a tridiagonal matrix representation 
of the algebra of the partially asymmetric exclusion process (PASEP)
lends itself to interpretation as the transfer matrix for weighted Motzkin lattice paths.
A continued fraction (``J-Fraction'')
representation of the lattice path generating function is particularly well
suited to discussing the PASEP, for which the paths have height dependent weights. We show that
this not only allows a succinct derivation of the normalisation and correlation lengths of the PASEP,
but also reveals how finite-dimensional representations of the PASEP algebra, valid only along special
lines in the phase diagram, relate to the general solution that requires an infinite-dimensional
representation.

\end{abstract}


\maketitle


\section{Introduction}

Although the asymmetric exclusion process (ASEP)---a model in which hard-core particles
hop in a preferred direction along a one-dimensional lattice---has been reinvented
in various different  
guises over the years, it is only relatively recently that
exact solutions for the steady state(s) of the model have been available.  The solution 
of the ASEP with open boundary conditions in \cite{DEHP} using a 
matrix product ansatz was a landmark in the study of driven 
diffusive systems.

As discussed in a recent review of the matrix product approach to solving for the steady state of nonequilibrium Markov processes \cite{MPrev}, there are a range of different methods for analysing the thermodynamic phase behaviour of the simplest versions of the ASEP. By contrast, more general models---collectively known as the partially asymmetric exclusion process (PASEP)---that admit particles to hop in both directions in the bulk, and even more generally to enter and exit at both left and right boundaries, have so far been studied only through a diagonalisation of the matrices appearing in the formalism \cite{Sas,BECE,Sas2}.  In this work, we extend a technique that previously admitted an extremely quick derivation of the ASEP phase behaviour under various updating schemes \cite{Us1,Us2} to these more general models.

The idea is to consider the behaviour of a ``grand-canonical partition function'' for the model. More precisely, we examine the generating function of the normalization of the nonequilibrium steady-state distribution over an ensemble of different lattice lengths whose mean is controlled by a fugacity. The thermodynamic phase behaviour can then be read off from the singularities of this generating function. Whilst obtaining this generating function is straightforward for the ASEP \cite{Us1,Us2}, a convenient closed form for the PASEP has remained elusive.

Our aim here is to demonstrate that a representation of the generating function that allows the thermodynamic phase behaviour to be determined with relative ease takes the form of an infinite continued fraction. This we arrive at through an interpretation of the PASEP normalization as the (equilibrium) partition function of lattice paths, which we discuss in Section~\ref{paths} after recalling the model definition and its basic properties. In Section~\ref{cfracrep} we show how to analyse the singularities embedded in the continued fraction representation. The results we obtain are, of course, equivalent to those obtained within other approaches \cite{Sas,BECE,Sas2}. However, given that continued fractions are not frequently encountered in statistical mechanical contexts, we feel there is some value in using the PASEP as an illustrative example of how to handle them.

We find that the analysis is intimately related to an approach based on finite-dimensional matrix representations \cite{mal,jaf}, exact along special lines in the phase diagram, and that the continued fraction shows how these particular solutions and the general solution are related. We further show that the continued-fraction approach extends to the most general version of the PASEP, solved in \cite{Sas2}, and that one can access both currents and correlation lengths through it. Finally, we return to the lattice path picture to elucidate the equilibrium counterpart of a nonequilibrium phase transition identified in \cite{BECE} that occurs when the bias on bulk hop rates opposes that imposed by the boundary conditions.

\section{Model Definition and Basic Properties}

The dynamics of the PASEP take place on a 
finite one-dimensional lattice with open boundaries.
In its simplest form, the microscopic dynamics of the PASEP are 
specified by four rates, one of which can be set to unity by an overall scaling.
For a rate $\lambda$ associated with a
particular event, the probability that the event happens in an infinitesimal
time interval $\Delta t$ is $\lambda \Delta t$. Moves that would
lead to two particles occupying a single lattice site
at any one time
are prohibited due to the hard-core repulsion between them.

\begin{figure}[b]
\begin{center}
\includegraphics[scale=0.9]{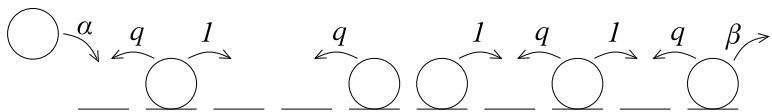}
\end{center}
\caption{\label{fig:PASEPrules}Typical particle configuration and
allowed moves in the PASEP model.}
\end{figure}

In the PASEP particles are inserted onto the left boundary site (when empty)  at a rate   $\alpha \,$
and removed from the right boundary site at a rate  $\beta \, $, see figure \ref{fig:PASEPrules}.
Once on the lattice a particle hops by one site to the right at rate $1 \,$
or by one site to the left  at a rate $q \, $ when sites are available (i.e.\ empty).
It is possible to expand this set of  moves to allow particles to  enter at the right at a rate $\delta$ and exit
at the left at a rate $\gamma$, while still retaining the solvability of the model \cite{Sas2}.

In all the models we consider  
we want to calculate  $Z$, which  normalises the statistical weight, $f(\Cee)$, 
of a lattice configuration, $\Cee$, in the steady state. This is given by
\begin{equation} \label{Zgen}
Z = \sum_{\Cee} f(\Cee)\;,
\end{equation}
so the normalized probability of being in state
$\Cee$ is $P(\Cee)=f(\Cee)/Z$.

The weights themselves are obtained through the stationarity condition
on the transition rates $W(\Cee \to \Cee^\prime)$,
\begin{equation}
\label{steadystate}
\sum_{\Cee^\prime \ne \Cee} \left[ f(\Cee^\prime) W(\Cee^\prime \to
\Cee) - f(\Cee) W(\Cee \to \Cee^\prime) \right] = 0
\;,
\end{equation}
where $W(\Cee \to \Cee^\prime)$ is the probability of making the transition 
from 
configuration $\Cee$ to $\Cee^\prime$ in a single timestep.
This is less restrictive than the detailed balance
condition for equilibrium states, which is obtained when the
sum in equation (\ref{steadystate}) vanishes term by term.

The solution of the ASEP in \cite{DEHP} and the PASEP
in \cite{Sas,BECE,Sas2} made use of a matrix product ansatz \cite{MPrev}.
In this the steady-state probability $P(\mathcal{C})$ 
of a configuration of particles $\mathcal{C}$ on a chain of length $N$
is represented by an ordered product of matrices $X_1 X_2 \ldots X_N$
where $X_i=D$ if site $i$ is occupied and $X_i=E$ if it is empty. 
We expect $P(\mathcal{C})$ to be a function of both the number and position of particles on the lattice, 
which suggests the choice of non-commuting objects, matrices,
for the ansatz.
To
obtain a scalar probability value from this matrix product
it is sandwiched between
two vectors $\bra{W}$ and $\ket{V}$:
\begin{equation}
\label{eqn:Pansatz}
P(\mathcal{C})=\frac{\braopket{W}{X_1 X_2 \ldots X_N}{V}}{Z_N} \;.
\end{equation}
The factor $Z_N$ is included to ensure that $P(\mathcal{C})$ is
properly normalised. 
This latter quantity plays the role
of a partition function in equilibrium problems
\begin{equation}
\label{eqn:Zdef}
Z_N=\braopket{W}{(D+E)^N}{V}=\braopket{W}{C^N}{V} \;,
\end{equation}
where we have defined $C=D+E$. Indeed, we shall see
in what follows
that $Z_N$ {\it is} the partition function for an equivalent
two-dimensional lattice path problem.

The algebraic properties
of the matrices $D$ and $E$ can be deduced from the master equation for the
dynamics of the ASEP, PASEP and various other related models \cite{MPrev}.  For
the variant of the PASEP discussed above, sufficient conditions for
equation (\ref{eqn:Pansatz}) to hold are
\begin{eqnarray}
\label{eqn:DEcommute}
DE-qED &=& D+E \;,\\
\label{eqn:EonW}
\alpha \bra{W}E &=& \bra{W} \;,\\
\label{eqn:DonV}
\beta D \ket{V} &=& \ket{V} \;.
\end{eqnarray}
These relations allow one to calculate $Z_N$ and other quantities
of physical interest by a range of methods, such as ``normal-ordering''
of the matrices, or through use of explicit representations \cite{MPrev}.

In this work, we focus on an approach based around the generating function
of $Z_N$, namely ${\cal Z}(z) = \sum_N Z_N z^N$, which can be thought of as
a ``grand-canonical'' normalization.  As is well known \cite{Wilf}, the large-$N$
form of the ``canonical'' normalization ($Z_N$) can be determined from the dominant
singularity $z_{cr}$ of ${\cal Z}(z)$. Typically, $Z_N \sim z_{cr}^{-N} N^{-\nu}$ where the exponent $\nu\ge0$ depends on the nature of the singularity. Then, by defining a
``reduced free energy'' $f$ via
\begin{equation}
f = - \lim_{N \to \infty} \frac{1}{N} \ln Z_N \;,
\end{equation}
we find $f = \ln z_{cr}$. Nonanalyticities in $f$ can then be associated with phase
transitions in the physical system \cite{Zeros,Us1,Us2,MPrev}.

For orientation, let us recall the results for the ASEP, which has $q=0$. The
canonical normalization can be shown by direct matrix reordering \cite{DEHP} to
be
\begin{equation}
\label{asepZ}
Z_N = \sum_{p=1}^{N} \frac{p(2N-1-p)!}{N!(N-p)!}  \frac{
(1/\beta)^{p+1} - (1/\alpha)^{p+1} }{ (1/\beta)-(1/\alpha) } \; .
\end{equation}
Performing the summation \cite{Us1} gives the grand canonical normalization
\begin{equation}
 {\cal Z} ( z ) = \frac{\alpha \beta }{ (\alpha - x (z ) ) ( \beta - x(z))} \; ,
\label{xaxb} 
\end{equation}
where $x(z)= (1 - \sqrt{1 - 4 z }) /2$\,.

\begin{figure}
\begin{center}
\includegraphics[scale=0.75]{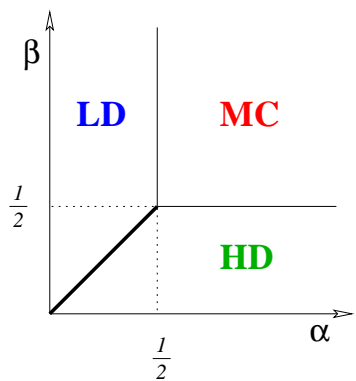}
\caption{\label{basicpd} The phase diagram of the ASEP. Here HD, LD and MC denote the high-density, low-density and maximal-current phases, respectively.}
\end{center}
\end{figure}

This function has a pole at $x(z)=\alpha$ when $\alpha<\frac{1}{2}$ and similarly at $x(z)=\beta$ when $\beta<\frac{1}{2}$. These correspond to $z_{cr}=\alpha(1-\alpha)$ and $z_{cr} = \beta(1-\beta)$. When neither of these poles contribute, all that remains is the square-root singularity at $z_{cr}=\frac{1}{4}$. This allows one to very quickly establish the behaviour of the reduced free energy as a function of $\alpha$ and $\beta$:
\begin{equation}
f = \left\{ \begin{array}{ll}
 \ln  \left[ \frac{1}{4} \right] & \mbox{for $\alpha, \; \beta> 1/2$} \\
 \ln \left[ \alpha ( 1  - \alpha ) \right] & \mbox{for $\beta>\alpha, \; \alpha<1/2$}\\
 \ln \left[ \beta ( 1 -  \beta ) \right]  & \mbox{for $\alpha > \beta , \; \beta<1/2$}
 \end{array} \right. \;. 
\end{equation}
It turns out that for the ASEP, $z_{cr}$ corresponds to the particle current, and $x(z_{cr})$ the bulk density in the thermodynamic limit. Hence, the phase diagram for the model, Figure~\ref{basicpd}, is quickly recovered using this generating-function (or grand-canonical) analysis. Further details of these methods as applied to the ASEP can be found in \cite{Us1,Us2,MPrev}.  In the remainder of this work, we show how to elicit the structure of the grand-canonical normalization of the PASEP, where direct summation of the canonical normalization, given explicitly in \cite{BECE}, does not lead to a compact expression like (\ref{xaxb}).

\section{Explicit Matrix Representation and Lattice Path Interpretation}
\label{paths}

A useful route to the grand-canonical normalization for the PASEP is via the generating function for an ensemble of lattice paths, which in turn can be read off from an explicit representation of the matrices and vectors appearing in the equations (\ref{eqn:DEcommute}), (\ref{eqn:EonW}) and (\ref{eqn:DonV}) that define the matrix algebra.  A number of representations are known, see \cite{DEHP,Sas,Sas2,MPrev}; the one that is of use here is that for which the vectors $\bra{W}$ and $\ket{V}$ have nonzero entries only in their first element:
\begin{equation}
\bra{W_{q}} = \bra{W} = h_0^{1/2} (1,0,0,\cdots ) \qquad
\ket{V_{q}} = \ket{V} = h_0^{1/2} (1,0,0,\cdots )^T \;,
\end{equation}
where $h_0$ is a constant to be given shortly.  One can verify that, with this choice of boundary vectors, the following tridiagonal representations of $D$ and $E$ satisfy 
(\ref{eqn:DEcommute}), (\ref{eqn:EonW}) and (\ref{eqn:DonV}):
\begin{eqnarray}
D_{q} &=& {1 \over 1 - q} \left(
\begin{array}{cccc}
1+ \tb    & \sqrt{c_1}    & 0             & \cdots\\
0     & 1 + \tb q  & \sqrt{c_2}    & {}\\
0               & 0     & 1 + \tb q^2  & \ddots\\
\vdots          & {}            & \ddots        & \ddots
\end{array}
\right) \;,
\nonumber \\
E_{q} &=& {1 \over 1 - q} \left(
\begin{array}{cccc}
1 + \ta    & 0    & 0             & \cdots\\
\sqrt{c_1}       & 1 + \ta q  &  0    & {}\\
0               & \sqrt{c_2}     & 1 + \ta q^2  & \ddots\\
\vdots          & {}            & \ddots        & \ddots
\end{array}
\right) \;.
\label{eqn:repde2}
\end{eqnarray}
The various parameters that appear are 
\begin{eqnarray}
\ta &=& { 1 - q \over \alpha} -1 \, , \\
\tb &=& { 1 - q \over \beta} -1 \, ,\\
c_n &=& (1 - q^n ) ( 1 - \ta \tb q^{n-1} ) \, ,\\
h_0 &=& \frac{1}{(\ta \tb;q)_\infty} = \sum_{n=0}^{\infty} \frac{(\ta \tb)^n}{(q;q)_n} = \braket{W}{V} \;,
\end{eqnarray}
in which we have used the standard notation for (shifted) $q$-factorials
\begin{eqnarray}
\label{eqn:qfacdef}
(a;q)_n &=& \prod_{j=0}^{n-1} (1-aq^j) \;,\nonumber \\
(a;q)_0 &=& 1 \;, \nonumber\\
(a,b,\ldots, c;q)_n &=& (a;q)_n (b;q)_n \ldots (c;q)_n \;.
\end{eqnarray}

There are various ways to arrive at an interpretation in terms of lattice paths from a matrix representation. One was suggested by Brak and Essam \cite{BrakEssam,BrakRitt}, who used the fact that $D + E = DE$ for the ASEP to interpret the $D$ and $E$ as odd-even and even-odd height transfer matrices separately. In later works \cite{Us1,Us2}, the path interpretation was inferred from the grand-canonical normalization once this had been obtained by another means. Here, we shall take the most direct approach, which is to associate a height $n \ge 0$ above the origin with the vector $\ket{n} = (0\; 0\; \cdots \;0\; 1\;0 \cdots )^T$ (i.e., $n$ is the number of zero entries that appear before the single nonzero entry). We then interpret $\bra{m} X \ket{n}$, where $X$ is some combination of $D$ and $E$ matrices, as the weight of paths connecting a point at height $n$ to another point at height $m$.

Of particular importance is the matrix $X=C^N$, which appears in Eq.~(\ref{eqn:Zdef}) for the normalization. From the above expressions we have that
\begin{equation}
\label{C}
C_{q}=D_{q} +E_{q} = {1 \over 1 - q} \left(
\begin{array}{cccc}
2+ \ta +\tb   & \sqrt{c_1}    & 0             & \cdots\\
 \sqrt{c_1}    & 2 + (\ta + \tb )q  & \sqrt{c_2}    & {}\\
0               &   \sqrt{c_2}   & 2 + (\ta + \tb) q^2  & \ddots\\
\vdots          & {}            & \ddots        & \ddots
\end{array}
\right) \;.
\end{equation}
The matrix element $\bra{m} C^N \ket{n}$ then gives the combined weight of paths that begin at height $m$, end at height $n$ and contain $N$ steps, each of which may raise or lower the height by one unit, or leave the height unchanged. Since the height is a nonnegative quantity, $n\ge0$, these paths may never descend beneath the origin. Additionally, since $\bra{W} \propto \bra{0}$ and $\ket{V} \propto \ket{0}$, the paths that contribute to $Z_N$ begin and end at the origin. Paths with these properties are known as \emph{Motzkin paths}.

\begin{figure}
\begin{center}
\includegraphics[width=0.45\linewidth]{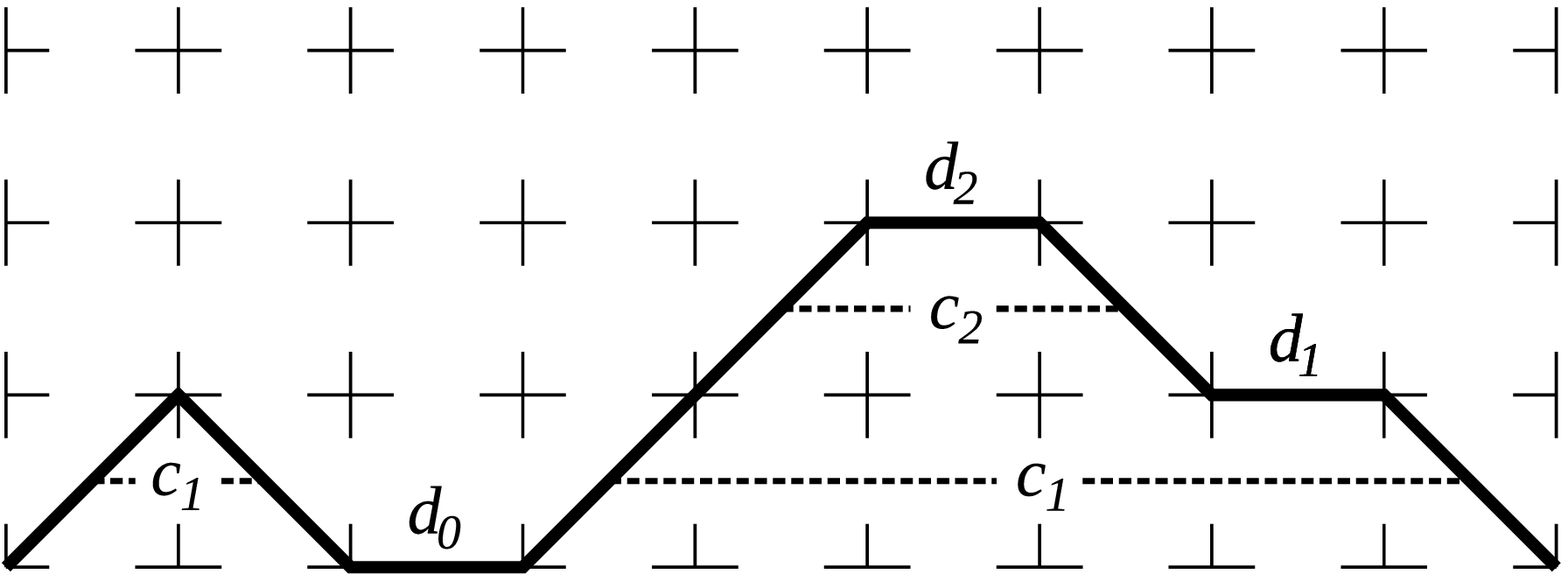}
\hspace{1em}
\raisebox{-4pt}{\includegraphics[width=0.45\linewidth]{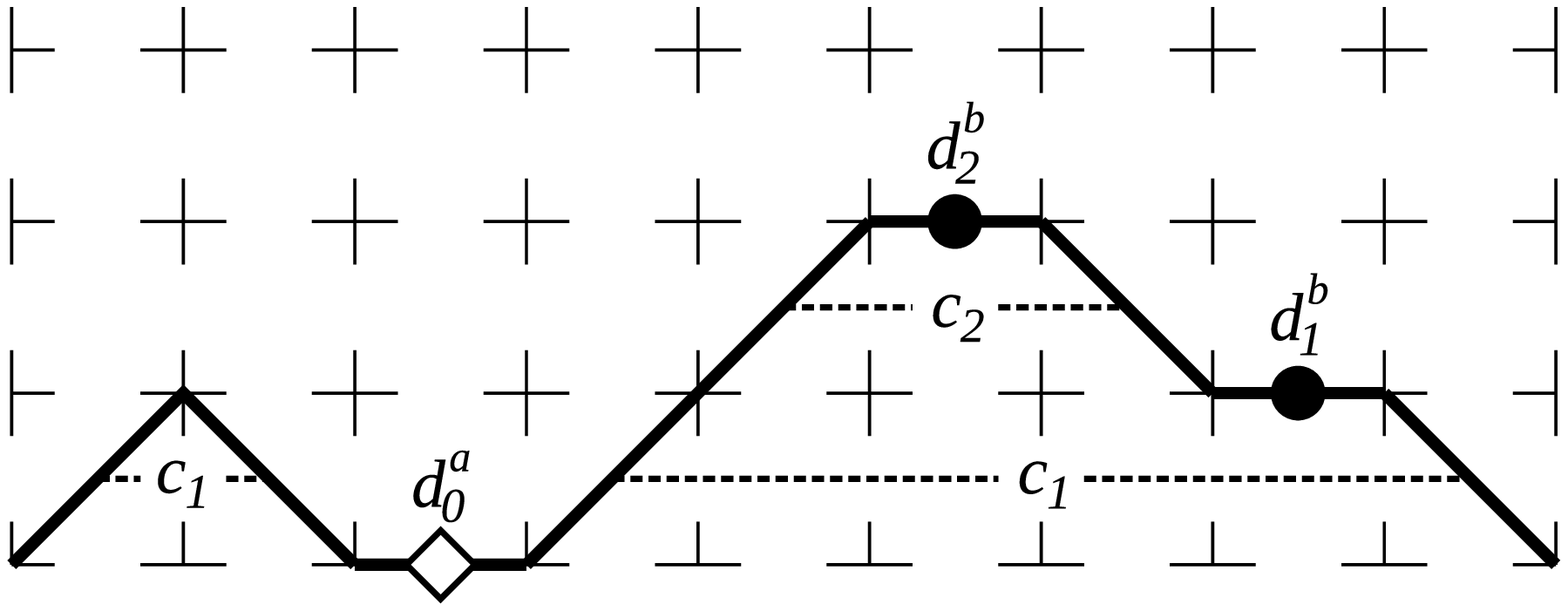}}
\caption{\label{motzkin-fig} Two Motzkin path transliterations of the tridiagonal matrix representation of the PASEP
matrices. In both cases, up-step down-step pairs contribute a weight $c_n$, where $n$ is the height above the origin
of the upper end of the steps. In the left-hand figure, there is a single type of horizontal step that contributes a
weight $d_n$, where $n$ is the height of the segment above the origin. The right-hand figure shows an equivalent
interpretation, in which there are two types of horizontal steps, shown with diamonds and circles at the their midpoints.
These contribute weights $d_n^{a}$ and $d_n^b$, respectively.}
\end{center}
\end{figure}

The weight of various path components can now be obtained by inspecting the form of $C_q$. For the path to begin and end at the origin, every up-step must be accompanied by a down-step; each up-step down-step pair connecting height $n-1$ to height $n$ contributes a weight $c_n$ to the path. Each horizontal step at height $n$ contributes a weight $d_n = 2+ (\ta + \tb) q^n$ to the path. An alternative interpretion has two types (or ``colours'') of horizontal path segments, one of which contributes a weight $d_n^{a} = 1 + \ta q^n$ and the other $d_n^{b}=1 + \tb q^n$. See figure~\ref{motzkin-fig}. To arrive at the canonical normalization $Z_N$, we sum over all paths of length $N$, and multiply by the factor $h_0 / (1-q)^N$. The grand-canonical normalization is constructed by summing paths of all lengths, weighting each segment by $z$, and finally multiplying by $h_0$. In the next section, we shall see an equivalent recursive construction which can be expressed as a continued fraction.

As a check of the path representation one can compare the expression for 
$Z_N$ which emerges from directly evaluating the matrix product expression for the stationary 
state, 
\begin{equation}
\label{eqn:Zsumrep:q<1}
Z_N=\braket{W}{V} \left( \frac{1}{1-q} \right)^{\!\!N}
\sum_{n=0}^N R_{N,n}(q) B_n(\tilde \alpha,\tilde \beta;q) \;,
\end{equation}
where
\begin{eqnarray}
\label{eqn:Rdef2}
{} \nonumber \\
R_{N,n}(q)=
\textstyle
\sum_{k=0}^{\halfint{N-n}} (-1)^k
\left[ \binom{2N}{N-n-2k} - \binom{2N}{N-n-2k-2} \right]
q^{\binom{k+1}{2}} \qbinom{n+k}{k}
\;,
\nonumber \\
{}
\end{eqnarray}

and
\begin{equation}
\label{eqn:Bdef}
B_n(\tilde \alpha, \tilde \beta;q)=\sum_{k=0}^{n} \qbinom{n}{k} \tilde \alpha^{n-k} \tilde \beta^k \;,
\end{equation}
with the weights which emerge from the contributing paths at low orders. In the above we have used the standard
notation for the $q$-binomial coefficient
\begin{equation}
\qbinom{n}{k} = \frac{(q;q)_n}{(q;q)_{n-k} (q;q)_k} \; .
\end{equation}
For instance, taking the simple case of $Z_2$ a direct calculation using the above formulae gives
\begin{equation}
\label{eqn:Z2}
Z_2 = { 5 - q + \ta^2 + \ta \tb + \tb^2 + q \ta \tb + 4 \ta + 4 \tb \over ( 1 - q)^2 }\;,	
\end{equation}
where we have dropped the overall normalization $\braket{W}{V}$.
The $(1-q)^2$ denominator disappears when this is written in terms of the original rate parameters $\alpha$ and $\beta$,
\begin{equation}
Z_2 = { \alpha \beta q+ \alpha^2+ \beta^2 \alpha+ \alpha \beta + \alpha^2 \beta + \beta^2 \over \alpha^2 \beta^2 } \; .
\end{equation}
The expression for $Z_2$ in (\ref{eqn:Z2}) can be seen to be the sum of weights for an up/down step pair from level $0$ to $1$,
given by $(1-q)(1 - \ta\tb)/ ( 1 -q)^2$ and all four possible combinations of two horizontal steps at level zero of either ``colour'', given by
$(2 + \ta + \tb)(2 + \ta + \tb)/(1-q)^2$ (where we have again  dropped the overall normalization $\braket{W}{V}$).
While the diagrammatics becomes increasingly complicated for larger $Z_N$, the principle remains the same.

\section{The Continued Fraction Representation of the Path Generating Function}
\label{cfracrep}

Despite the availability of an exact expression for the canonical normalization, Eq.~(\ref{eqn:Zsumrep:q<1}) from \cite{BECE}, we have not been able to find a convenient expression for its grand-canonical counterpart due to the
$q$-dependence of the PASEP weights. A more fruitful route is to represent the lattice-path generating function
as a continued fraction, a procedure first expounded by Flajolet \cite{Flaj}. As we now show, this representation
can be read off more-or-less directly from the matrix representation (\ref{C}), and the form that emerges is particularly well adapted to discussion of the PASEP.

First, for convenience, we subsume the prefactor $1/(1-q)$ appearing in (\ref{C}) into the parameters
\begin{eqnarray}
\tilde d_n = {2 + ( \ta + \tb ) \, q^n \over 1 -q } \;,\nonumber \\
\tilde c_n = {  (1 - q^n ) ( 1 - \ta \tb \, q^{n-1} ) \over ( 1 - q)^2} 
\label{cn}
\;,
\end{eqnarray}
so that then
\begin{equation}
C_{q}=  \left(
\begin{array}{cccc}
\tilde{d}_0   & \sqrt{\tilde c_1}    & 0             & \cdots\\
 \sqrt{\tilde c_1}    & \tilde{d}_1 & \sqrt{\tilde c_2}    & {}\\
0               &   \sqrt{\tilde c_2}   & \tilde{d}_2 & \ddots\\
\vdots          & {}            & \ddots        & \ddots
\end{array}
\right) \;.
\end{equation}
In the lattice-path language, this means that horizontal steps at height $n$ are weighted by $\tilde d_n$
and up-down step pairs between heights $n$ and $n+1$ are weighted by $\tilde c_n$.

Let now ${\cal M}_n(z)$ be the generating function of weighted Motzkin paths that start and end at height $n$, never go below this height, and have their lengths $N\ge 0$ counted by powers of $z$. That is, the coefficient of $z^N$ in ${\cal M}_n(z)$ is the weight of such paths of length $N$. The grand-canonical normalization for the PASEP is then given by ${\cal Z}(z) = h_0 {\cal M}_0(z)$. Let us suppose that ${\cal M}_{n+1}(z)$ is known for some $n \ge 0$. Then, we can construct ${\cal M}_n(z)$ by concatenating contiguous components of two types: (i) sequences of horizontal segments of arbitrary (possibly zero) length at height $n$; and (ii) Motzkin paths starting at height $n+1$ enclosed by an up-down pair. Denoting these components schematically as $\underbar{\hspace{2.5ex}}$ and $\diagup {\cal M}_{n+1} \diagdown$ respectively, we can write the recursion
\begin{eqnarray}
 {\cal M}_n &=& \underbar{\hspace{2.5ex}} + \underbar{\hspace{2.5ex}} \diagup {\cal M}_{n+1} \diagdown  \underbar{\hspace{2.5ex}} + \underbar{\hspace{2.5ex}} \diagup {\cal M}_{n+1} \diagdown  \underbar{\hspace{2.5ex}} \diagup {\cal M}_{n+1} \diagdown  \underbar{\hspace{2.5ex}} + \cdots \\
&=& \underbar{\hspace{2.5ex}} \left( 1 + \left[ \diagup {\cal M}_{n+1} \diagdown  \underbar{\hspace{2.5ex}} \right] + 
\left[ \diagup {\cal M}_{n+1} \diagdown  \underbar{\hspace{2.5ex}} \right]^2 + \cdots \right) \\
&=& \frac{ \underbar{\hspace{2.5ex}} }{ 1 - \diagup {\cal M}_{n+1} \diagdown  \underbar{\hspace{2.5ex}} } \;.
\end{eqnarray}
The generating function for a (possibly empty) sequence of horizontal segments, each weighted by $\tilde{d}_n$, is simply $(1 - \tilde{d}_n z)^{-1}$. An up-down pair $\diagup \cdots \diagdown$ from height $n$ to $n+1$ contributes the weight $\tilde{c}_{n+1} z^2$. We thus arrive at the generating-function recursion
\begin{equation}
 {\cal M}_n(z) = \frac{(1 - \tilde{d}_n z)^{-1}}{1 - \tilde{c}_n z^2 {\cal M}_{n+1}(z)(1 - \tilde{d}_n z)^{-1}}
= \frac{1}{1 - \tilde{d}_n z - \tilde{c}_{n+1} z^2 {\cal M}_{n+1}(z)} \;.
\end{equation}
Starting at $n=0$ and iterating, we find that the generating function ${\cal Z}{( \ta, \tb, q, z )}$ for Motzkin paths 
of arbitrary length (and hence the grand-canonical PASEP normalization)
is given by the infinite continued fraction
\begin{equation}
 {\cal Z} ( \ta, \tb, q, z ) = \frac{1}{\displaystyle 1 - \tilde d_0 z -
\frac{\tilde c_1 z^2}{\displaystyle 1 - \tilde d_1 z - 
\frac{\tilde c_2 z^2}{\displaystyle 1 - \tilde d_2 z - 
\frac{\tilde c_3 z^2}{\displaystyle \ldots
}}}} \; \; \; ,
\label{fracq}
\end{equation}
where we have dropped the factor $h_0$ since this does not contribute to any physical quantities.
Such a continued fraction containing both $z$ and $z^2$ terms is usually denoted a \emph{Jacobi continued fraction}, or ``J-fraction''
for short \cite{Wall}.

Before considering the case of general $q$ let us take $q=0$ and see how the expression for the grand canonical normalization, Eq.~(\ref{xaxb}), is recovered. We have
\begin{equation}
 {\cal Z} ( \ta, \tb, 0, z ) = \frac{1}{\displaystyle 1 - \tilde d_0 z -
\frac{\tilde c_1 z^2}{\displaystyle 1 - 2 z - 
\frac{z^2}{\displaystyle 1 - 2 z - 
\frac{z^2}{\displaystyle 1 - 2 z -
\frac{z^2}{\displaystyle \ldots
}}}}} \;\;\;,
\label{fracq0}
\end{equation}
where $\tilde d_0 = 1/ \alpha + 1/ \beta$ and $\tilde c_1 = 1 / \alpha + 1 / \beta - 1/ ( \alpha \beta ) = \kappa^2$ when $q=0$\,. Note that in this case, the continued fraction is periodic after the first level. That is, in the above notation,
\begin{equation}
 {\cal M}_n(z) = \frac{1}{1 - 2z - z^2 {\cal M}_{n+1}(z)} \quad \forall n \ge 1 \;.
\end{equation}
Hence, we must have that ${\cal M}_1(z) = {\cal M}_2(z) = \cdots$ and hence
\begin{equation}
 {\cal M}_1(z) \left[ 1 - 2z - z^2 {\cal M}_1(z) \right] = 1
\end{equation}
or
\begin{equation}
{\cal M}_1(z) = { 1 - 2 z - \sqrt{1 - 4 z } \over 2 z^2 } \;,
\end{equation}
which is the generating function familiar from many Catalan counting problems. For $n=0$, we have
 \begin{equation}
 {\cal Z} ( \ta, \tb, 0, z ) = {\cal M}_0(z) = \frac{1} { 1 - \tilde d_0 z - \tilde c_1 z^2 {\cal M}_1(z)} \;,
\label{fracq0a}
\end{equation}
which coincides with (\ref{xaxb}) when both are expanded and rationalized.

At $q=0$ the luxury of being able to sum the continued fraction
 to get equation (\ref{fracq0a}) makes the phase structure, 
which (as previously discussed) is determined by the singularities 
of ${\cal Z}(\ta, \tb, 0, z )$ in $z$, immediately apparent.  If we are {\it not} able to easily sum explicitly the continued fraction, as is the case 
for the PASEP, we can use more indirect methods to determine the singularities. We focus here on the ``forward-bias'' regime, $q<1$, in which the continued fraction has a finite radius of convergence in the complex-$z$ plane. In the reverse-bias regime, $q>1$, the continued fraction is unconditionally divergent, a fact we will interpret physically in Section~\ref{pathtrans}. 

We first appeal to Worpitzsky's theorem on the convergence of continued fractions
\cite{Wall,JvR} which states that a continued fraction of the form
\begin{equation}
\frac{1}{\displaystyle 1 +
\frac{a_2}{\displaystyle 1 + 
\frac{a_3}{\displaystyle 1  + 
\frac{a_4}{\displaystyle \ldots
}}}} 
\nonumber
\end{equation}
converges if the partial numerators $a_p$ satisfy 
\begin{equation}
| a_p | < 1/4, \; p = 2, 3, 4, \dots \;.
\end{equation}
For ${\cal Z} (z)$ given by (\ref{fracq0}) this translates to
a radius of convergence $z_{cr}$ given by
\begin{equation}
{4 \tilde c_n z_{cr}^2 \over ( 1 - \tilde d_{ n -1 } z_{cr} ) ( 1 - \tilde d_n z_{cr}) } = 1 \; \, \, \forall n
\label{worp2}
\end{equation} 
and shows that 
\begin{equation}
z_{cr} \to (1 - q)/4
\label{zw}
\end{equation}
as $n \to \infty$. To decide if this is the
dominant singularity, one must also divine the location of any poles in the complex-$z$ plane from the continued fraction (\ref{fracq}).

The strategy is to examine the $n^{\rm th}$ \emph{convergent} of the continued fraction, that is, the expression obtained by truncating the continued fraction at the $n^{\rm th}$ level (counting from zero). Denoting this as $K_n$, we have that
\begin{eqnarray}
K_0 &=& \frac{1}{1 - \tilde{d}_0 z} \; , \\
K_1 &=& \frac{1}{1 - \tilde{d}_0 z - \frac{\tilde{c}_1 z^2}{1 - \tilde{d}_1 z} } \; ,
\end{eqnarray}
and so on. We observe that the continued fraction (\ref{fracq0}) is given \emph{exactly} by the convergent $K_n$ if $\tilde{c}_{n+1}=0$. Inspection of (\ref{cn}) reveals that this occurs on the special line in the phase diagram given by
\begin{equation}
\label{abcond}
\tilde{\alpha} \tilde{\beta} = q^{-n} \;.
\end{equation}
We observe that on such special lines, the matrix $C_q$ decomposes into two blocks: the first is $(n+1)$-dimensional, whilst the second does not contribute to the normalisation because only the first elements of the boundary vectors are nonzero. Finite-dimensional representations of the matrix algebra for the PASEP were used prior to the advent of the full solution to study the model along these lines, and to conjecture the phase behaviour elsewhere \cite{mal,jaf}.

To locate the poles of $K_n$ in the complex-$z$ plane, we use the fact \cite{Special} that the reciprocal of each convergent $K_n$ can be expressed as
\begin{equation}
(K_n)^{-1} = \frac{A_n}{B_n}
\end{equation}
where $A_n$ and $B_n$ both satisfy the same three-term recurrence
\begin{eqnarray}
\label{Anrr}
A_n(z) &=& (1 - \tilde{d}_n z) A_{n-1}(z) - \tilde{c}_n z^2 A_{n-2}(z) \; , \\
B_n(z) &=& (1 - \tilde{d}_n z) B_{n-1}(z) - \tilde{c}_n z^2 B_{n-2}(z) \; ,
\end{eqnarray}
but with slightly different initial conditions: $A_{-2}=B_{-1}=0$ and $A_{-1}=B_{0}=1$.  Hence, $K_n$ has poles that coincide with the zeros of $A_n(z)$. These relations for $A_n(z)$ and $B_n(z)$ hold for arbitrary choices of $\ta$ and $\tb$. Let us now define the function ${\cal A}_n(z)$ that is given by $A_n(z)$ for the particular choice $\tb = 1/(q^n \ta)$. That is, the zeros of ${\cal A}_n(z)$ give the poles of the PASEP grand-canonical generating function along the line in the phase diagram along which the $(n+1)$-dimensional matrix representation is exact.

As a consequence of the recurrence (\ref{Anrr}) the $A_n(z)$ are closely related to the al-Salam-Chihara polynomials \cite{AC}
and may be written explicitly as
\begin{equation}
\label{Ansum}
A_n( z) =	\frac{(\ta \tb;q)_{n+1} z^{n+1}}{ (1-q)^{n+1} \ta^{n+1}} \; 
\sum_{k=0}^{n+1} \frac{(q^{-(n+1)};q)_k (\ta \, e^{i \theta};q)_k (\ta \, e^{-i
\theta};q)_k}{(\ta \tb;q)_k (q;q)_k} q^k
\end{equation}
where  $\cos( \theta) = (1-q) / ( 2 z) -1$. If we take  $\tb = 1/(q^n \ta)$ to obtain ${\cal A}_n(z)$ all the terms in the sum  in (\ref{Ansum}) will be zero apart from the
$(n+1)th$ since they  contain a $(1- \ta \tb q^n)$ factor in the numerator, which is cancelled
only in the last term. Using 
\begin{equation}
	(\ta e^{i \theta};q)_{n+1} (\ta e^{-i\theta};q)_{n+1}
= \prod_{j=0}^{n} ( 1 - 2 \ta q^j \cos(\theta) + \ta^2 q^{2j} )
\end{equation}
we thus find
\begin{equation}
{\cal A}_n(z) = \prod_{k=0}^{n} \left(1 - \frac{z}{z_k} \right)
\end{equation}
where
\begin{equation}
z_k = \frac{1-q}{ \bigg( 1 + \ta q^k \bigg) \bigg( 1 + \frac{1}{\ta q^k} \bigg) } \;.
\end{equation}
Equally, we could have taken $\ta = 1/(q^n \tb)$ and obtained the same result with $\ta \leftrightarrow \tb$. These roots correspond to the reciprocal of the eigenvalues of the finite-dimensional matrices $C_q$ examined in \cite{mal,jaf} (note, however, that the representation used in those works differs from that used here). This fact can be understood by making the ansatz $A_n(z) = z^n | C_q^{(n)} - \frac{1}{z} |$, where $C_q^{(n)}$ is the submatrix formed by the first $n$ rows and columns of $C_q$. After substituting into (\ref{Anrr}), one obtains the expression that results from a co-factor expansion along the bottom row in the determinant $|C_q^{n} - \lambda|$, as long as $\lambda$ is identified with $\frac{1}{z}$.

We see, then, that the sequence of generating functions ${\cal A}_n(z)$ has a truly remarkable property: all poles of ${\cal A}_n(z)$ are poles of ${\cal A}_m(z)$ where $m > n$. In principle, a completely different set of singularities could have been obtained after truncating at a deeper level.  Furthermore, when $q<1$, $|z_n| < |z_m|$ if $n<m$, and hence all convergents have the same dominant pole, $z_0$.  Writing this in terms of the  original rates $\alpha$ and $\beta$ gives
\begin{equation}
z_{cr} = \frac{1- q}{2 + \ta + 1 / \ta} = { \alpha ( 1 - q - \alpha ) \over 1 - q } 
\label{zaq}
\end{equation}
or
\begin{equation}
z_{cr} = \frac{1-q}{2 + \tb + 1 / \tb} =  { \beta ( 1 - q - \beta ) \over 1 - q } \;.
\label{zbq} 
\end{equation}
Taking now the number of levels (or, equivalently, the dimensionality of the matrix representation) to infinity, we find that the dominant singularity will always be the smallest of (\ref{zw}), (\ref{zaq}) or (\ref{zbq}). This observation allows us to recover the reduced free energies for the PASEP previously calculated in \cite{BECE,Sas}, and identify the regions of the phase diagram within which they apply:
\begin{eqnarray}
\label{eq:fsdef}
f &=& \ln \left[ \frac{1-q}{4} \right] \;, \qquad{\rm for}\; \alpha, \; \beta> (1 - q)/2\;,  \nonumber \\
f &=& \ln \left[ \frac{ \alpha ( 1 - q - \alpha )}{1-q} \right] \;, \qquad{\rm for} \; \beta>\alpha, \; \alpha<(1-q)/2 \; ,\ \label{eq:frsc<1} \nonumber \\
f &=& \ln \left[ \frac{ \beta ( 1 - q - \beta ) } { 1 - q} \right]\;, \qquad{\rm for}\; \alpha > \beta , \; \beta<(1-q)/2 \;. \label{eq:frsc>1}
\end{eqnarray}
As for the ASEP, this phase structure comprises high- and low-density phase and a maximal current phase. The key differences lie in the $q$-dependent position of the second-order
transition line, $\alpha = \beta = (1 - q) /2$, and in the subextensive corrections to the free energy, which manifest themselves in the density profile at finite distances from the boundaries \cite{Sas3}.

As noted above, this analysis of the PASEP phase behaviour is similar to that based on finite-dimensional matrix representations \cite{mal,jaf}.  The main additional benefit of the present approach is that the relationship between the general expression for the normalisation, given through the non-terminating continued fraction (\ref{fracq}), and the versions that apply only on special lines is clearer.  For example, the singularity that arises from non-convergence of the continued fraction, and which governs the maximal-current behaviour, is not detectable by inspection of the finite dimensional matrices. Furthermore, some light is shed on the ``continuity arguments'' expounded in \cite{mal,jaf} that lead one to extend physical properties found along special lines to regions of the phase diagram, in that each additional level of the continued fraction yields a singularity that is subdominant to all of those present at the previous level. We remark that the fact that the asymptotically dominant singularity is present even in the one-dimensional representation is suggestive of an explanation as to why the phase diagram and currents obtained from mean-field theory coincide with the exact results.

\section{Lattice Paths for Non-Zero ``Wrong Direction'' Removal and Injection Rates $\gamma$ and $\delta$}

As we noted in the introduction
it is possible to generalize the PASEP boundary conditions to allow 
``wrong direction'' injection and removal rates.
In addition to injection at a rate $\alpha$
at the left boundary particles may also be removed there  at a rate $\gamma$. Similarly,
in addition to removal at a rate $\beta$ at the right boundary particles may also be injected there at a rate $\delta$. 

A tridiagonal representation of $D$ and $E$ with these generalized boundary conditions still exists for the case of non-zero $\gamma$ and $\delta$ \cite{Sas2} and which can
be written as
\begin{eqnarray}
\hat D_{q} &=& {1 \over 1 - q} \left(
\begin{array}{cccc}
1 + d_0^{\natural}    & d_0^{\sharp}    & 0             & \cdots\\
d_0^{\flat}     & 1 + d_1^{\natural}     & d_1^{\sharp}    & {}\\
0               & d_1^{\flat}     & 1 + d_2^{\natural}   & \ddots\\
\vdots          & {}            & \ddots        & \ddots
\end{array}
\right) \;,
\nonumber \\
\hat E_{q} &=& { 1 \over 1 - q } \left(
\begin{array}{cccc}
1 + e_0^{\natural}    & e_0^{\sharp}    & 0             & \cdots\\
e_0^{\flat}       & 1 + e_1^{\natural}   &  e_1^{\sharp}    & {}\\
0               & e_1^{\flat}     & 1 + e_2^{\natural}   & \ddots\\
\vdots          & {}            & \ddots        & \ddots
\end{array}
\right) \;,
\label{eqn:repde2a}
\end{eqnarray}
\begin{eqnarray}
\bra{\tilde W_{q}} = h_0^{1/2} (1,0,0,\cdots ) \;, \qquad
\ket{\tilde V_{q}} = h_0^{1/2} (1,0,0,\cdots )^T \;,
\label{eqn:repWV2a}
\end{eqnarray}
where the expressions for $d_n^{\sharp}$, $d_n^{\natural}$, $e_n^{\natural}$, $e_n^{\flat}$ (a notation borrowed from \cite{Sas2}) and $h_0$ are given in the appendix.

We can define $\hat C_q = \hat D_q + \hat E_q$ in an analogous manner to the earlier discussion which leads to a generating function of the form 
\begin{equation}
 {\cal Z} ( a, b, c, d, q, z ) = \frac{1}{\displaystyle 1 - \hat d_0 z -
\frac{\hat c_1 z^2}{\displaystyle 1 - \hat d_1 z - 
\frac{\hat c_2 z^2}{\displaystyle 1 - \hat d_2 z - 
\frac{\hat c_3 z^2}{\displaystyle \ldots
}}}} \;\;\;, 
\end{equation} 
where the coefficients are now
\begin{eqnarray}
\hat d_n = {2 + d_n^{\natural}+ e_n^{\natural} \over 1 -q } \;, \nonumber \\
\hat c_{n+1} =  {(d_n^{\sharp} + e_n^{\sharp}) 
( d_n^{\flat} + e_n^{\flat})  
\over ( 1 - q)^2 } \;.
\end{eqnarray}
Even with these more complicated coefficients
the singularity structure of the generating function 
is identical to the $\gamma=\delta=0$ case when $q<1$. 
We can still extract  poles 
in order to discern the high- and low-density phases.
Explicitly,
\begin{eqnarray}
{} \nonumber \\
\hat{c}_{n+1} &=&
\textstyle
\frac{(1-q^{n-1}abcd)(1-q^{n+1})
(1-q^nab)(1-q^nac)(1-q^nad)(1-q^nbc)(1-q^nbd)(1-q^ncd)}
{(1-q^{2n-1}abcd)(1-q^{2n}abcd)^2(1-q^{2n+1}abcd) (1 - q)^2} \nonumber \\
{} 
\end{eqnarray}
(in which the parameters $a$, $b$, $c$ and $d$ are also given in the appendix)
so that we can pick up the various poles depending on the zeros of
the numerator. For non-zero $\gamma$ and $\delta$ Askey-Wilson polynomials \cite{AW} play the 
role of the al-Salam-Chihara polynomials so in this case the denominators of the convergents
are given by
\begin{eqnarray}
\label{Ansum2}
{} \nonumber \\
\textstyle
A_n( z)=\frac{(ab, ac , ad;q)_{n+1} z^{n+1}}{ (1-q)^{n+1} a^{n+1}} \; 
\sum_{k=0}^{n+1} \, \frac{(q^{-(n+1)};q)_k (q^n a b c d) (a \, e^{i \theta};q)_k (a \, e^{-i
\theta};q)_k}{(a b;q)_k (a c;q)_k (a d;q)_k(q;q)_k} \, q^k \nonumber \\
{}
\end{eqnarray}
where  $\cos( \theta) = (1-q) / ( 2 z) -1$ as before.
If we now choose $q^n a b =1$, for example, to enforce $\hat{c}_{n+1}=0$ we find that only the final
term in the sum contributes again, resulting in the dominant poles
\begin{equation}
z_{cr} = \frac{1 - q}{2 + a + 1 /a}
\end{equation}
or
\begin{equation}
z_{cr} = \frac{1 - q }{2 + b + 1 /b}
\end{equation}
depending on the relative size of $a$ and $b$. The maximal current phase is still observed in a similar manner to the $\gamma=\delta=0$ case via Worpitzky's theorem but a major difference  is that
the reverse bias phase is absent when $q>1$, since the particles can now escape at both ends. 
Algebraically this is manifested in the reflection symmetry relating the parameters 
for $q<1$ and $q>1$ when $\gamma$ and $\delta$ are non-zero:
\begin{equation}
a, b, c, d, q \rightarrow b^{-1}, a^{-1}, d^{-1}, c^{-1}, q^{-1}  \; .
\end{equation}
We remark that this continued-fraction representation provides a very quick route to the identification of the dominant singularities in the generating function, and therewith the extensive part of the free energy.

\section{Correlation Lengths}

We have so far  restricted our discussion to the normalization itself but it is also possible to extract
the correlation lengths, which determine the various sub-phases within the PASEP phase diagram, in a rather 
straightforward manner. We have seen that the leading singularities of the grand-canonical normalization $\cal{Z}$ determine the 
phase structure of the PASEP and that these can be extracted from the continued fraction representation either directly
as poles when the fraction terminates (the high- and low-density phases), or using Worpitzsky's theorem (the maximal current phase).

For a chain of length $N$ the one- and two-point density correlation functions for the PASEP may be obtained in terms 
of the matrices $D_q$ and $C_q$ as 
\begin{eqnarray}
\la \tau_i\ra  &=&
\frac{1}{Z_N}\la W\vert C_q^{i-1} D_q C_q^{N-i}\vert V\ra \;, \nonumber \\
\la \tau_i\tau_j\ra  &=&
\frac{1}{Z_N}\la W\vert C_q^{i-1}D_q C_q^{j-i-1} D_q C_q^{N-j}\vert V\ra \;.
\end{eqnarray}
We have already evaluated the thermodynamic limit  $N\to\infty$
of $Z_{N}$ by picking up the leading singularity in the partial fraction representation. If we switch back to the
transfer matrix picture of $C_q$, the leading singularity of the generating function $z_{cr}$ is just the inverse of the largest eigenvalue of $C_q$,
$\lambda_{cr} = 1 / z_{cr}$,
which dominates the normalization as
\begin{eqnarray}
\la W\vert C_q^N\vert V\ra &=& \sum_\lambda \la W\vert C_q^N\vert \lambda \ra\la \lambda \vert V\ra  \nonumber \\
&=& \sum_\lambda  \la W\vert \lambda \ra \lambda ^N\la \lambda \vert V\ra  \nonumber \\
&{}& \simeq \mathrm{const.} \, \lambda_{cr}^N \;.
\end{eqnarray}
For the two-point function, there is a  similar expansion  \cite{Uchi}
\begin{eqnarray}
&\langle W\vert C_q^{i-1}D_q C_q^{j-i-1}D_q
C_q^{N-j}\vert V\rangle \nonumber \\
&\simeq
\lambda_{cr}^{N-2} \la W\vert \lambda_{cr}\ra 
\la \lambda_{cr} \vert D_q\vert \lambda_{cr}\ra 
\la \lambda_{cr} \vert D_q\vert \lambda_{cr}\ra 
\la \lambda_{cr} \vert V\ra \nonumber\\
&\quad +\lambda_{cr}^{N-2} 
\left(\frac{\lambda_{sub}}{\lambda_{cr}}\right)^{i-1}
\la W\vert \lambda_{sub}\ra 
\la \lambda_{sub} \vert D_q\vert \lambda_{cr}\ra 
\la \lambda_{cr} \vert D_q\vert \lambda_{cr}\ra 
\la \lambda_{cr} \vert V\ra\\
&\quad +\lambda_{cr}^{N-2} 
\left(\frac{\lambda_{sub}}{\lambda_{cr}}\right)^{j-i-1}
\la W\vert \lambda_{cr}\ra 
\la \lambda_{cr} \vert D_q\vert \lambda_{sub}\ra 
\la \lambda_{sub} \vert D_q\vert \lambda_{cr}\ra 
\la \lambda_{cr} \vert V\ra \nonumber \\
&\quad +\lambda_{cr}^{N-2} 
\left(\frac{\lambda_{sub}}{\lambda_{cr}}\right)^{N-j}
\la W\vert \lambda_{cr}\ra 
\la \lambda_{cr} \vert D_q\vert \lambda_{cr}\ra 
\la \lambda_{cr} \vert D_q\vert \lambda_{sub}\ra 
\la \lambda_{sub} \vert V\ra \;, \nonumber
\end{eqnarray}
where the second largest eigenvalue $\lambda_{sub}$ is the inverse of the sub-dominant singularity $z_{sub}$
in the generating function ${\cal Z}$.
From this it is clear that the ratio $\lambda_{sub}/\lambda_{cr}$, or alternatively
$z_{cr} / z_{sub}$,
determines the correlation length.

If we consider first the high- and low-density phases 
the sub-dominant singularity is separated from the dominant one and both can be read off directly from 
the continued fraction representation of ${\cal Z}$.
In this case we descend two levels in  the continued fraction in order to extract both the 
dominant and sub-dominant singularities, which means that we impose $\tilde c_{2}=0$, i.e., $\tilde \beta  = 1 / (\tilde \alpha q )$
or $\tilde \alpha=1/(\tilde \beta q)$ depending on the values of the parameters. The generating function then truncates to
  \begin{eqnarray}
 {\cal Z}_2 ( \tilde \alpha, \tilde \beta, q, z ) &=& \frac{1}{\displaystyle 1 - \tilde d_0 z -
\frac{\tilde c_1 z^2}{\displaystyle 1 - \tilde d_1 z}
} \nonumber \\
&=& \frac{1 -\tilde d_{1} z}{1 - ( \tilde d_{0}+ \tilde d_{1}) z + ( \tilde d_{0} \tilde d_{1}- \tilde c_{1})z^{2}} \;,
\end{eqnarray}
which allows us to extract both the dominant singularity $z_{cr}$ and subdominant singularity
$z_{sub}$ for various ranges of the parameters $\tilde \alpha, \tilde \beta, q$. 
As we have already noted in extracting 
the normalization from the continued fraction, descending to deeper levels still picks up
$z_{cr}$ and $z_{sub}$ as the dominant and sub-dominant pole at every deeper level, in addition to the further sub-leading singularities. 

\begin{figure}
\begin{center}
\includegraphics[scale=0.4]{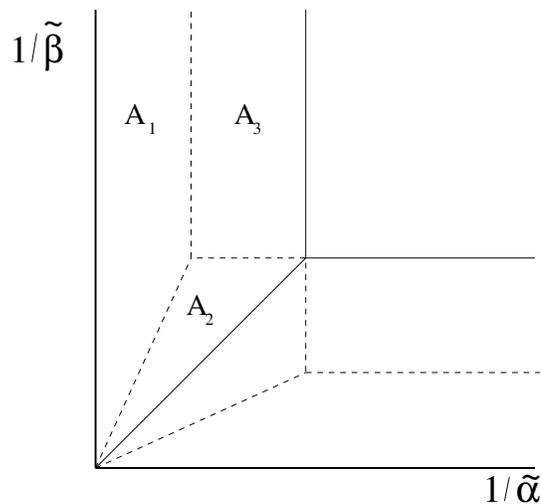}
\caption{\label{phasediagram} The various sub-phases ${\rm A}_1, {\rm A}_2, \dots$ of the PASEP
defined by the behaviour of the correlation length are delineated.
}
\end{center}
\end{figure}

This in turn allows us to classify various
sub-phases within the high- and low-density phases which have different behaviours of the correlation length.
For example, the low-density phase may be divided into the
three sub-phases found in \cite{Uchi}:
\\
\noindent
$({\rm A}_1) \; \; \; \tilde \alpha q >1,\ \tilde \alpha q > \tilde \beta$:
\begin{eqnarray}
z_{cr}=\frac{1-q}{(1+\tilde \alpha)(1+{\tilde \alpha}^{-1})} \;,\qquad
z_{sub}= \frac{1-q}{(1+{\tilde \alpha } q)(1+(\tilde \alpha q)^{-1})} \;;
\end{eqnarray}
$({\rm A}_2) \; \; \; \tilde \alpha > \tilde \beta > \tilde \alpha q, \ \tilde \beta >1$:
\begin{eqnarray}
z_{cr}=\frac{1-q}{(1+ \tilde \alpha)(1+{\tilde \alpha}^{-1})} \;,\qquad
z_{sub}=\frac{1-q}{(1+ \tilde \beta )(1+{\tilde \beta}^{-1})} \;;
\end{eqnarray}
$ ({\rm A}_3) \; \;   \;                     \tilde \alpha > 1 > \tilde \alpha q,\  \tilde \beta < 1$:
\begin{eqnarray}
z_{cr}=\frac{1-q}{(1+\tilde \alpha)(1+{\tilde \alpha}^{-1})} \;,\qquad
z_{sub}=\frac{1-q}{4} \;.
\end{eqnarray}
The $z_{sub}$ values which play a role in the $A_1$ and $A_2$ sub-phases are the (sub-dominant) poles of the truncated ${\cal Z}_2 (\tilde \alpha, \tilde \beta,q,z)$, whereas
in the $A_3$ sub-phase it is the singularity determined using Worpitzsky's theorem for the full, untruncated ${\cal Z}$.
The high-density phase  splits in a similar fashion with $\tilde \alpha $ and $ \tilde \beta$ interchanged in the various expressions
for $z_{cr}$ and $z_{sub}$ above. 
In all these sub-phases the correlation functions decay exponentially and the correlation length is 
given by the log of the ratio of the poles
 $\xi=\left(\ln \left(z_{sub}/z_{cr}\right)\right)^{-1}$. 
The phase structure deduced in this manner by considering the
singularities of ${\cal Z}$ agrees, as it should, with that obtained
from considering the integral representation of $Z_N$ in \cite{Uchi}
and is shown in figure~\ref{phasediagram}.

\section{(Motzkin) Path Transitions}
\label{pathtrans}

In our earlier work on Dyck path representations of the ASEP normalization it was
found that the high- and low-density phases of the ASEP were reflected
in bound phases for the lattice paths, whereas the maximal current phase
corresponded to an unbound phase.  The second-order transition lines
of the ASEP corresponded to unbinding transition lines for one of the two lattice paths 
which described the ASEP normalization and the first-order transition between the high- and low-density phases of the ASEP corresponded to a cooperative transition involving both paths.

The situation  for the 
$q<1$ phase structure of the PASEP is 
analogous. 
In the region of the phase diagram corresponding to the high- and low-density phases 
of the PASEP predominantly one type
of horizontal step is bound to the $x$-axis, reflecting
the majority of particles or holes in the high- and low-density phases of the PASEP.
In the maximal current phase the entropy dominates and paths are unbound. For instance,
in the low-density $A$ phases, indicated in figure~\ref{phasediagram}, the horizontal on-axis steps
weighted by $1 + \tilde \alpha $ will make a greater contribution to
the partition function than those weighted by
$1 + \tilde \beta$ and so will dominate the sum. The reverse is true in the high-density phase
where the steps weighted by $1 + \tilde \beta$ dominate.

The non-trivial terms in the weights of steps at height $n$ are proportional
to $q^n$ so these steps do not contribute to the energy as $N \to \infty$ when $q<1$, but only entropically. It is therefore not surprising that for $q<1$ the PASEP phase diagram is essentially identical
to the ASEP, with only the position of the transition lines changing as $q$ is varied. It is interesting 
to inquire about the nature of the paths contributing to the partition function for  $q > 1$. The expression
for $Z_N$ in this reverse bias phase was calculated in \cite{BECE},
\begin{equation}
\label{eqn:Z:q>1}
Z_N \sim A(\ta, \tb;q) \, (\invq \ta \tb,1/ \ta \tb;\invq)_\infty
     \left( \frac{\sqrt{\ta \tb}}{q-1} \right)^{\!\!N} q^{\frac{1}{4}N^2} \;,
\end{equation}
where
\begin{equation}
\label{eqn:Adef}
A(\ta,\tb;q) = \sqrt{\frac{\pi}{\ln q}} \exp \left[ M(q)+\frac{(\ln
    \tb / \ta)^2}{4 \ln q} \right] \;,
\end{equation}
and
\begin{equation}
\label{eqn:M:approx}
M(q) \simeq -\sum_{k=1}^\infty \frac{1}{k} \frac{1}{q^k-1} \; .
\end{equation}
The most interesting feature of (\ref{eqn:Z:q>1}) is that one now sees an area-like behaviour
emerging, whereby $Z_N \sim q^{\frac{1}{4}N^2}$. 

To see how such two-dimensional behaviour emerges from a model of one-dimensional paths
when $q>1$ consider a tent-shaped path contributing to $Z_N$ with $N/2$ upward steps followed by
$N/2$ downward steps. This has  a weight 
\begin{equation}
{ (\ta \tb ; q)_{N/2} (q;q)_{N/2} \over ( q - 1)^N} \;,
\end{equation}
which has a leading term of the form 
\begin{equation}
\label{eqn:leading}
{(\ta \tb - 1 ) (\ta \tb)^{{N \over 2} -1 } q^{N^2 \over 4} \over (q - 1 )^N } \; .
\end{equation}
The reverse bias expression for $Z_N$ in (\ref{eqn:Z:q>1}) displays exactly this behaviour. With a slight rewriting of the second factor,
\begin{equation}
\label{eqn:Z:q>1a}
Z_N \sim A(\ta, \tb;q) \, (\invq \ta \tb,1/ q \ta \tb;\invq)_\infty \left({\ta \tb - 1 \over \ta \tb} \right)
     \left( \frac{\sqrt{\ta \tb}}{q-1} \right)^{\!\!N} q^{\frac{1}{4}N^2} \;,
\end{equation}
which matches the behaviour of (\ref{eqn:leading}) coming from a ``tent'' path. Heuristically this 
makes sense since paths which gain in altitude as quickly as possible will donate the highest powers
of $q$ to $Z_N$ and dominate when $q>1$. 

This change in behaviour at $q>1$ to an area-like scaling is thus interpreted in the 
lattice path picture as an inflation transition, where the fluctuating path becomes the boundary of an inflated vesicle (with 
particular edge weights favouring a speedy ascent) pinned at its ends to the
horizontal axis. The matrices $D_q$ and $E_q$ initially denoted the presence or absence of particles on a particular site in the original matrix product 
ansatz. We can  associate an up-step and one colour of horizontal step in a path with, say, the presence of a particle in the original PASEP
steady state. Looking at the form of the leading contribution to $Z_N$ when $q>1$, namely $N/2$ up-steps followed by $N/2$ down-steps, we can see that this
corresponds to a half-filled lattice, which has also been observed in simulations of the reverse bias phase and heuristically derived using particle/hole symmetry arguments
\cite{BECE}.

The form of equation (\ref{eqn:Z:q>1}) is reminiscent of the partition function for inflated lattice vesicles with a fixed perimeter discussed in \cite{Prell}.
If one models a two-dimensional vesicle with a convex polygon of perimeter $2 N$ on a square lattice then the partition function is given by
\begin{equation}
Z_N(q) \sim  { ( 1 + O ( \rho^N)) \over ( q^{-1} ; q^{-1})_{\infty}^4} \sum_{k =1}^{N-1} q^{k ( N - k)} \;,
\end{equation}
where each component square making up the polygon has a weight $q$ and $\rho <1$. The summation arises from rectangles and may be extended
to run from $\pm \infty$ with an error of $O(q^{-{N^2 \over 4}})$ and the prefactor can be thought of as arising from nibbling off the corners of 
the rectangles with Ferrers diagrams as described in \cite{Prell}. In the case of the PASEP the expression for $Z_N$ when $q>1$ is suggestive of a similar interpretation; the
leading ``tent'' term is decorated with nibbled edges from the fluctuations.

\section{The Case $q=1$}

The expression for $Z_N$ may also be evaluated in the case of the symmetric simple exclusion process (SSEP) when $q=1$.
If we introduce some more standard notation, this time  for $q$-numbers,
\begin{equation}
[\, n \, ]_q = \frac{\; 1 - q^n}{1 - q} = 1 + q + \dots q^{n-1} \; ,
\end{equation}
(with $[\, n \, ]_q = 0$ for $n \le 0$) we can rewrite the Motzkin-path weights as 
\begin{eqnarray}
\tilde d_n &=& 2 [\, n \, ]_q  + \left( {1 \over \alpha} + { 1 \over \beta } \right) q^n \;, \nonumber \\
\tilde c_n &=&  [\, n \, ]_q  \left( [\, n -1 \, ]_q
+ \left( {1 \over \alpha} + { 1 \over \beta } \right) q^{n-1}
- { ( 1 - q ) q^{n-1} \over \alpha \beta } \right) \;,
\end{eqnarray}
or, splitting the horizontal weights into two types of steps again in order to make the correspondence
with bicoloured Motzkin paths explicit,
\begin{eqnarray}
\tilde d_{n,1} &=& [\, n \, ]_q  + {q^n\over \alpha} \;, \nonumber \\
\tilde d_{n,2} &=& [\, n \, ]_q  + {q^n \over \beta}  \;.
\end{eqnarray}
The $q$-integers become normal integers when $q=1$, so the weights at $q=1$
are given by
\begin{eqnarray}
\tilde d_{n,1} &=& n   + {1\over \alpha} \;, \nonumber \\
\tilde d_{n,2} &=&  n  + {1 \over \beta} \;, \nonumber \\
\tilde c_n     &=&  n    \left( n -1 +  {1 \over \alpha} + { 1 \over \beta }   \right) \; ,
\end{eqnarray}
which simplify even further when $\alpha=\beta=1$ to $n+1$ and $n ( n+1)$. This gives
\begin{equation}
 {\cal Z} ( 1 , 1 , 1, z ) = \frac{1}{\displaystyle 1 - 2 z -
\frac{2 z^2}{\displaystyle 1 -  4 z - 
\frac{2 \times 3 z^2}{\displaystyle 1 -  6  z - 
\frac{ 3 \times 4 z^2}{\displaystyle \ldots
}}}} \; \; \;, 
\label{fracqq}
\end{equation}
which can be seen to be a continued fraction expansion for the divergent power series
\begin{equation}
{\cal Z} = \sum_{N=0}^{\infty} (N+1) ! \, z^N \;,	
\end{equation}
so $Z_N = (N+1)!$ in this case.
A similar expansion exists for general $\alpha$ and $\beta$ and gives $Z_N$ as the ratio of two
Gamma functions,
\begin{equation}
Z_N = {\Gamma ( \lambda + N + 1) \over \Gamma  ( \lambda + 1 ) }
\end{equation}	
where $\lambda = {1\over \alpha} + {1 \over \beta} -1$. 
This agrees with a direct calculation in \cite{Sas2}. 
Using Stirling's approximation for the Gamma functions,
and remembering that $\lambda$ is of $O(1)$ we see that
$\ln Z_N \sim N \ln N$ so $Z_N \sim e^{ N \ln N}$ at $q=1$, intermediate between the linear behaviour for $q<1$ and
the area law behaviour when $q>1$.  

\section{The Case $q^{n+1}=1$}

The continued fraction representation of ${\cal Z}$ will also terminate when $q^{n+1}=1$.
These values of $q$ have recently proved of interest for the Bethe Ansatz approach to the open $XXZ$ spin chain with non-diagonal boundary conditions \cite{Nepo1,Fab1,Fab2}. The Hamiltonian of the $XXZ$ spin chain is equivalent (up to a unitary transformation) to the transfer matrix of the PASEP. If we denote
the expression for $A_n(z)$ when $q^{n+1} = 1$ by $\tilde{\cal A}_n (z)$ the sum in (\ref{Ansum})  truncates to two terms 
\begin{equation}
	\tilde{\cal A}_n (z) = {\frac{z^{n+1}}{(1-q)^{n+1}\ta^{n+1}}} \left( (\ta \tb ; q)_{n+1} - (\ta e^{i \theta} ; q)_{n+1} (\ta e^{-i \theta}; q )_{n+1} \right)
\end{equation}
where we have used a limiting procedure 
\begin{equation}
\lim_{\epsilon \to 0} \;  \frac{(q^{-n+1};q)_{n+1}}{(q;q)_{n+1}}= \lim_{\epsilon \to 0} \; (-1)^{n+1} q^{-(n+1)(n+2)/2} = -1
\end{equation}
with $q= \exp ( \epsilon + 2 \pi  i m / ( n + 1 ))$  to handle the indeterminate final term  \cite{Spirid}. Noting that $(\ta e^{i \theta},q)_{n+1} = (1 - \ta^{n+1} e^{i ( n+1) \theta})$ when $q^{n+1}=1$
this can be written as
\begin{equation}
	\tilde{\cal A}_n (z) = {\frac{z^{n+1}}{(1-q)^{n+1}}} \left( e^{i (n+1)\theta} + e^{-i(n+1)\theta} - (\ta^{n+1}+ \tb^{n+1} ) \right) 
\end{equation}
and we find that the zeros of $\tilde{\cal A}_n (z)$ are given by 
\begin{equation}
z_k = { 1 - q  \over (1  + r q^k)(1  +  \frac{1} { r q^k})}
\end{equation}
where $r$ is the root with smaller argument of
\begin{equation}
r^{n+1} = {(\ta^{n+1}+ \tb^{n+1} ) \over 2} + \sqrt{{(\ta^{n+1}+ \tb^{n+1} )^2 \over 4} - 1 }
\end{equation}
or
\begin{equation}
r^{-(n+1)} = {(\ta^{n+1}+ \tb^{n+1} ) \over 2} - \sqrt{{(\ta^{n+1}+ \tb^{n+1} )^2 \over 4} - 1 }.
\end{equation}
It is interesting that the overall structure of the roots remain similar to that observed when $\ta = 1/ ( q^n \tb )$, but one no longer has a dominant pole appearing at first order and remaining.

\section{Conclusions}

We have shown that the matrix $C_q = D_q + E_q$ in a particular
tridiagonal  
representation of the PASEP algebra can be interpreted as the transfer
matrix for weighted Motzkin paths, with two colours of horizontal
steps. 
Writing  the generating function for  
these paths as a continued fraction allowed a  succinct derivation of the 
thermodynamic limit of the normalization of the 
PASEP in its various phases. In particular it allowed calculations 
without explicitly summing the generating function
in closed form, which was possible (or at any rate, easy to perform) only in the $q=0$ case. 
Consideration of the sub-leading singularities in the continued
fraction also allowed the determination of the
correlation length in the high- and low-density phases.

A further interesting feature of the continued fraction respresentation of the generating function was that 
it made clear how the  finite-dimensional representations of the PASEP algebra, valid only along special
lines in the phase diagram, related to the general, infinite-dimensional
solution via truncation. The unusual structure of the poles of the convergents of the continued fraction
along these special lines was highlighted,
with lower order poles remaining present as the order of the convergent increased and the dominant pole
(for the high- and low-density phases) already being present at zeroth order. The presence of the 
maximal current phase, on the other hand, was deduced from the continued fraction respresentation by using Worpitzky's
theorem.
 
The phase transitions of the PASEP can be identified
with transitions in the lattice path model. When $q<1$
there are similar unbinding transitions to those seen in
the Dyck path model for the ASEP, with bound states
corresponding to the high- and low-density phases
of the PASEP and an unbound phase corresponding to
the maximal current phase. For the reverse bias phase of the PASEP when $q>1$
the lattice path model was seen to be in an inflated phase and the leading
contribution to the lattice path partition function was identified with a half-filled state
in the PASEP.

Interestingly, bicoloured Motzkin paths play a major role
in various combinatorial bijections 
and the ASEP and PASEP have recently been investigated 
for their combinatorial interest \cite{Sylvie1,Sylvie2,Sylvie3,Gilles1,Gilles2}.
One common theme has been 
the appearance of various well-known combinatorial weights
as special cases
for the (P)ASEP normalization which can be related to known results
in the enumeration of permutations and other combinatorial objects. The approach in \cite{Sylvie3} in particular
is rather similar to that espoused here, namely going directly to a matrix representation for $D$ and $E$ and seeking
a combinatorial interpretation (although the principal focus in \cite{Sylvie3} was on a different representation
and permutation tableaux).

It would be an interesting exercise to relate the phase transitions discussed here for Motzkin paths and the
related phase structure of the PASEP to conformational transitions in other (weighted) combinatorial objects such 
as parallelogram polyominoes, binary trees  and the permutation tableau of \cite{Sylvie3}. It would also be worthwhile
to relate the lattice path picture to discussions of the large deviation functional of the (P)ASEP, such as that in
\cite{largeD}, where the appearance of a combination of a Brownian excursion and  a Brownian walk in the discussion is
strongly suggestive of a polyomino bijection from the Motzkin paths here.


\section{Acknowledgements}

This work was partially supported by the 
EU RTN-Network `ENRAGE': {\em Random Geometry
and Random Matrices: From Quantum Gravity to Econophysics\/} under grant
No.~MRTN-CT-2004-005616.

\appendix

\section{Parameters Appearing in the Tridiagonal Matrix Representation for $\gamma,\delta \ne 0$}

When all four boundary rates, $\alpha$, $\beta$, $\gamma$ and $\delta$ are nonzero, the tridiagonal $D$ and $E$ matrices (\ref{eqn:repde2a}) contain a number of parameters. These are
\begin{eqnarray*}
d_n^\natural &=&
\frac{q^{n-1}}{(1-q^{2n-2}abcd)(1-q^{2n}abcd)} \\
&&\times[
bd(a+c)+(b+d)q-abcd(b+d)q^{n-1}-\{ bd(a+c)+abcd(b+d)\} q^n \\
&&-bd(a+c)q^{n+1}+ab^2 cd^2(a+c) q^{2n-1}+abcd(b+d)q^{2n} ] \;, \\
e_n^\natural &=&
\frac{q^{n-1}}{(1-q^{2n-2}abcd)(1-q^{2n}abcd)} \\
&&\times[
ac(b+d)+(a+c)q-abcd(a+c)q^{n-1}-\{ ac(b+d)+abcd(a+c)\} q^n \\
&&-ac(b+d)q^{n+1}+a^2 bc^2 d(b+d) q^{2n-1}+abcd(a+c)q^{2n} ] \;, 
\end{eqnarray*}
\begin{eqnarray*}
&&d_n^\sharp =
\frac{1}{1-q^nac}\mathcal{A}_n \;,\qquad
e_n^\sharp =
-\frac{q^nac}{1-q^nac}\mathcal{A}_n \;,  \\
&&d_n^\flat =
-\frac{q^nbd}{1-q^nbd}\mathcal{A}_n \;,\qquad
e_n^\flat =
\frac{1}{1-q^nbd}\mathcal{A}_n \;,
\end{eqnarray*}
which involve the further parameters
\begin{eqnarray*}
a &=& \frac{1}{2 \alpha} \left[ ( 1 - q - \alpha + \gamma) + \sqrt{ ( 1 - q - \alpha + \gamma)^2 + 4 \alpha \gamma} \right] \;,  \\
b &=& \frac{1}{2 \beta} \left[ ( 1 - q - \beta + \delta) + \sqrt{ ( 1 - q - \beta + \delta)^2 + 4 \beta \delta} \right] \;,  \\
c &=& \frac{1}{2 \alpha} \left[ ( 1 - q - \alpha + \gamma) - \sqrt{ ( 1 - q - \alpha + \gamma)^2 + 4 \alpha \gamma} \right] \;,  \\
d &=& \frac{1}{2 \beta} \left[ ( 1 - q - \beta + \delta) - \sqrt{ ( 1 - q - \beta + \delta)^2 + 4 \beta \delta} \right] \\
\mathcal{A}_n &=&
\textstyle
\left[
\frac{(1-q^{n-1}abcd)(1-q^{n+1})
(1-q^nab)(1-q^nac)(1-q^nad)(1-q^nbc)(1-q^nbd)(1-q^ncd)}
{(1-q^{2n-1}abcd)(1-q^{2n}abcd)^2(1-q^{2n+1}abcd)}
\right]^{1/2} \;.
\end{eqnarray*}
Finally, the constant appearing in (\ref{eqn:repWV2a}) is
\begin{equation}
h_0 = { ( a b c d ; q )_{\infty} \over ( q , ab , ac , ad , bc, bd, cd ; q)_{\infty} } \;.
\end{equation}

\vspace{1cm}



\end{document}